\documentclass[prl,reprint]{revtex4-1}
\usepackage{graphicx}
\usepackage{verbatim}
\usepackage{bbold}

\usepackage{amsmath}

\begin{document}

\title{The optimal bound of quantum erasure with limited means}
\author{Filippo M. Miatto$^{1}$, Kevin Pich\'e$^{1}$, Thomas Brougham$^{2}$, and Robert W.~Boyd$^{1,2,3}$}
\affiliation{$^1$Dept.~of Physics, University of Ottawa, Ottawa, Canada}
\affiliation{$^2$School of Physics and Astronomy, University of Glasgow, Glasgow (UK)}
\affiliation{$^3$Institute of Optics, University of Rochester, Rochester, USA}
\date{\today}

\begin{abstract}
In practical applications of quantum information science, quantum systems can have non-negligible interactions with the environment, and this generally degrades the power of quantum protocols as it introduces noise. Counteracting this by appropriately measuring the environment (and therefore projecting its state) would require access all the necessary degrees of freedom, which in practice can be far too hard to achieve. 
To better understand one's limitations, we calculate the upper bound of optimal quantum erasure (i.e.  the highest recoverable visibility, or ``coherence''), when erasure is realistically limited to an accessible subspace of the whole environment. In the particular case of a two-dimensional accessible environment, the bound is given by the sub-fidelity of two particular states of the \emph{inaccessible} environment, which opens a new window into understanding the connection between correlated systems. We also provide an analytical solution for a three-dimensional accessible environment. This result provides also an interesting operational interpretation of sub-fidelity.
We end with a statistical analysis of the expected visibility of an optimally erased random state and we find that \emph{i}) if one picks a random pure state of 2 qubits, there is an optimal measurement that allows one to distill a 1-qubit state with almost 90\% visibility and \emph{ii}) if one picks a random pure state of 2 qubits in an inaccessible environment, there is an optimal measurement that allows one to distill a 1-qubit state with almost twice its initial visibility.
\end{abstract}
\maketitle

\section{Introduction}
Complementarity is one of the jewels of quantum mechanics. It was first introduced by Bohr \cite{Bohr1}, as a consequence of the uncertainty principle. However, it took several decades to establish that its origin is really due to quantum correlations \cite{Scully1991,Wiseman1995,Durr1998}. The principle of complementarity gained its modern form through the works of several authors \cite{WZ1972,GreenbergerYasin,Scully1989,Sanders1989,Durr1998}.
In particular, Englert gave a very lucid exposition of the connection between complementarity and the working principles of a two-way interferometer \cite{Englert1996}. As the state of a quantum system in a two-way interferometer can be described as a simple qubit, an effective way of studying complementarity is through our familiarity with the Bloch sphere. This intuition is the key to also understand quantum erasure, i.e. the ability of restoring coherence in a system by appropriately projecting another system that is correlated to it and that is preventing the occurrence of interference \cite{Scully1982,Kwiat1992,Herzog1995}.

We now describe the situation that we are considering and the concepts that we will adopt. A state that lives in a 2-dimensional Hilbert space can be described in the language of quantum information as a qubit. Due to the possible embedding of this Hilbert space in a larger one (which in our choice of language represents the ``environment''), correlations of both classical and quantum nature can exist between the two. In this situation, the reduced state of the qubit is not pure, i.e. it has a certain degree of mixedness. 
Complete knowledge of a quantum state implies that such state is pure and in fact, a possible strategy to restore coherence is to gather the necessary knowledge from the environment by way of a suitable measurement. When the environment is measured, the qubit is projected on the state that is relative to the outcome of the measurement, and for an optimal choice of measurement, the projected state can be pure. However, there can be different choices of optimal measurements, that give rise to different final results. In particular, if we fix a preferred basis in the 2-dimensional space of the qubit, we can pick a measurement that maximizes the degree of superposition of the two basis vectors or one that maximizes the amplitude of one basis vector over the other. These two measurement choices are both optimal in the sense that they maximize some criterion, and we will refer to them as \emph{quantum erasure} and the \emph{which-alternative} measurements, respectively \cite{Englert2000}.
In the Bloch sphere picture, where the preferred basis is represented by the two poles, a quantum erasure measurement on the environment projects the qubit states towards the equator, while the which-alternative measurement projects the qubit states towards one of the poles.
\begin{figure}[!t]
\begin{center}
\includegraphics[width=0.5\columnwidth]{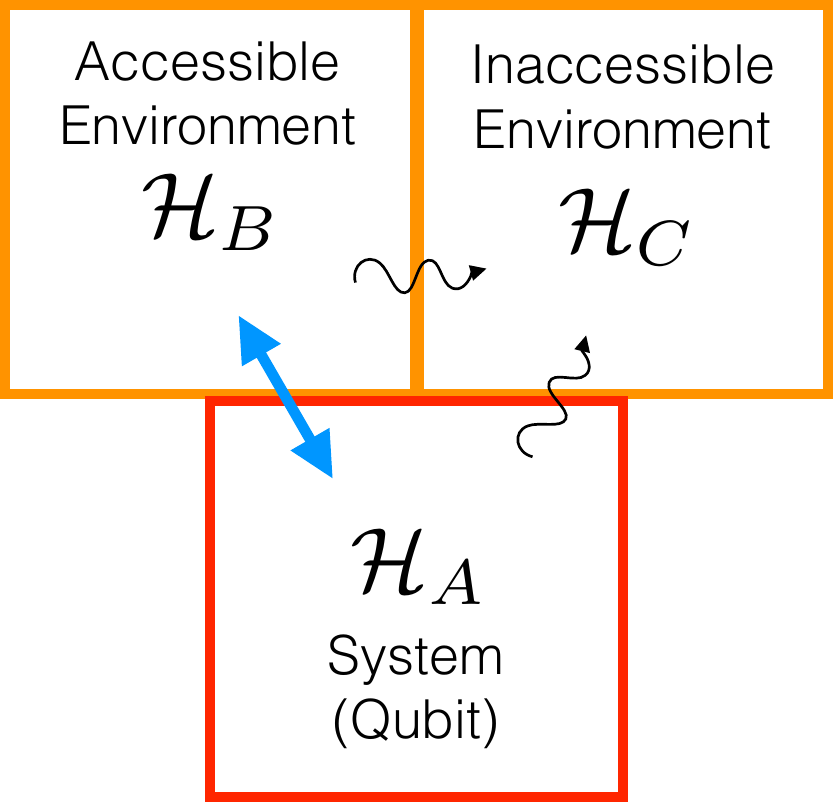}
\caption{\label{fig1} The environment (upper boxes) is split into an accessible subspace $\mathcal H_B$ and an inaccessible subspace $\mathcal H_C$ of arbitrary dimension. The blue arrows remind us that the three systems are generally correlated. Our ability of restoring coherence in $\mathcal H_A$ by acting on $\mathcal H_B$ depends on the sub-fidelity of the conditional states in $\mathcal H_C$.}
\end{center}
\end{figure}

\begin{figure}[!t]
\begin{center}
\includegraphics[width=0.75\columnwidth]{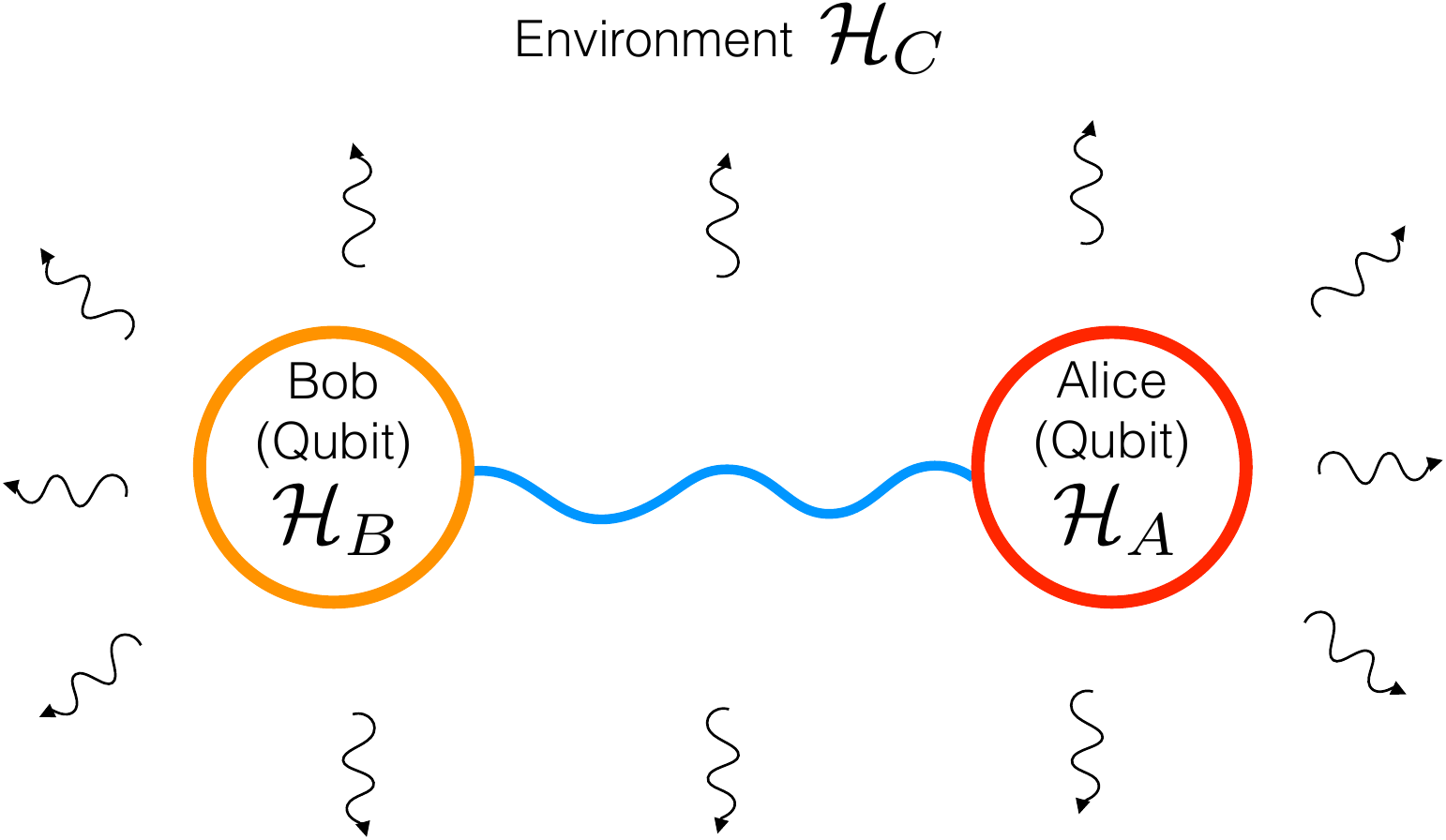}
\caption{\label{alicebob} An alternative scenario is that of remote state preparation of system $B$, where some information about the joint $AB$ system could leak to an inaccessible environment.}
\end{center}
\end{figure}

From this introduction it would seem rather feasible to control the qubit by way of measurements on the environment, but this operation is limited by two factors. The first is of physical nature: we can control the qubit to a degree that depends on how strong the correlations with the environment are. As a limiting case, if the two systems were independent we would have no control over the qubit by manipulating the environment. The second is of technical nature: in order to have the highest degree of control allowed by the strength of the correlations, one would need to be able to perform the \emph{desired} measurement, i.e. to project on the desired axes of the Hilbert space of the environment, which implies the ability of manipulating all the necessary degrees of freedom. This can be very hard to achieve, and in the case of an environment with too many degrees of freedom it is practically impossible.
In this work, we want to quantify the degree of control in the limiting case of minimal access to the environment: we consider a 1-qubit state embedded in an arbitrary-dimensional environment, and we split the Hilbert space of the environment into an accessible part $\mathcal H_B$ and an inaccessible one $\mathcal H_C$ (as we will be referring to these three systems rather often, we will indicate them by $A$, $B$ and $C$ throughout the paper). Then, we quantify the highest average visibility that one can retrieve on the qubit $A$ by appropriately measuring the accessible environment $B$. Geometrically, it is the largest \emph{average} distance of all the outcomes from the line connecting the poles of the Bloch sphere, see Fig.~\ref{fig1}. We find a surprising analytical answer to this problem, in terms of the sub-fidelity of conditional states of the inaccessible environment $C$, Eq.~\eqref{result}.

\section{Quantifying visibility and predictability}
In our analysis we do not allow for selective measurements, the reason is that selective measurements (i.e. postselection) allow one to achieve a considerable flexibility at the expense of probability of success, whereas we are interested in ``one-shot'' measurements, which cannot rely on postselection. These would occur, for instance, when a measurement takes place too far into a quantum algorithm and it would be too inconvenient to start over, or if two parties cannot communicate, as in a remote state preparation scheme. In general, if we had complete access to the environment, the strength of correlations between the qubit and the environment would be the only limitation on our ability to indirectly prepare the qubit. However, if we could perform a measurement over and over until the desired outcome occurs (say, if we had an inexpensive source of identically prepared states), we would be able to eventually prepare the qubit regardless of the strength of the correlations (as long as they are not zero). On the other hand, if selective measurements were not allowed, the states that the qubit could reach after a measurement of the environment would be restricted by the strength of the correlations. 

Regarding as ``environment'' the whole set of quantum systems that are correlated to the qubit (so that the state of qubit+environment is pure), we now prove that a successful measurement of a rank-1 projector in the whole environment space projects the qubit in a conditional pure state which can reach any point in the Bloch sphere (the price to pay is a probability of success which in general is less than 1): 
start with the joint qubit+environment state
\begin{align}
\label{fullstate}
|\rho\rangle=\sqrt{p_0}|0,e_0\rangle+\sqrt{p_1}e^{i\phi}|1,e_1\rangle,
\end{align}
where the qubit is in the computational basis.
If $|\rho\rangle$ is non-separable, it must hold that $|e_0\rangle\neq|e_1\rangle$, so it is possible to write $|e_0\rangle=\alpha|e_1\rangle+\beta|e_1^\bot\rangle$ for an appropriate choice of $|e_1^\bot\rangle$ orthogonal to $|e_1\rangle$ which implies $|e_1\rangle=\alpha^*|e_0\rangle+\beta e^{i\theta}|e_0^\bot\rangle$, with $|e_0^\bot\rangle\neq|e_1^\bot\rangle$. Consider then a successful measurement of the environment in the state $a|e_0^\bot\rangle+b|e_1^\bot\rangle$. This projects the qubit in the conditional state $\frac{\beta}{\sqrt{P_s}}(\sqrt{p_0}b^*|0\rangle+\sqrt{p_1}a^* e^{i(\theta+\phi)}|1\rangle)$ with a success probability $P_s=(p_1|b|^2+p_2|a|^2)|\beta|^2$. As $|0\rangle$ and $|1\rangle$ are orthogonal and as  $a$ and $b$ can be chosen freely, one can reach any pure state on the surface of the Bloch sphere. An immediate generalization allows one to conclude that using elements of a probability operator measure (POM) (also known as positive operator-valued measure, POVM), one can reach any state also in the interior of the Bloch sphere, and the proof is done.

\begin{figure}[t!]
\begin{center}
\includegraphics[width=0.7\columnwidth]{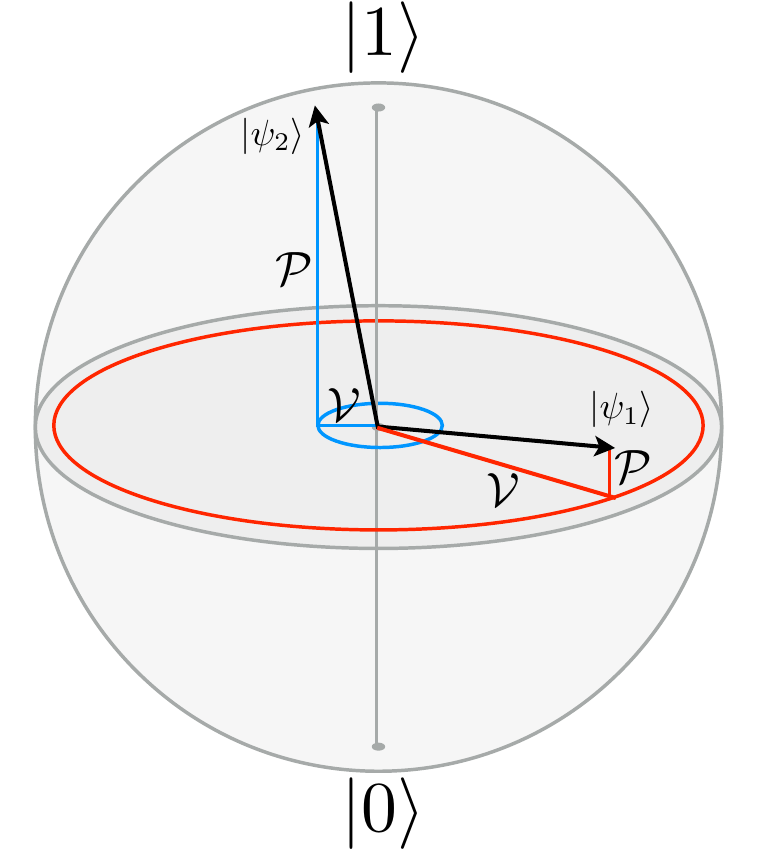}
\caption{\label{definitions}The visibility $\mathcal V$ can be understood as the distance from the North-South line and it indicates the degree of superposition of the two alternatives (here $|0\rangle$ and $|1\rangle$) and the predictability $\mathcal P$ as the distance from the equatorial plane. All states (here two examples of pure states) must satisfy the relation $\mathcal P^2+\mathcal V^2\leq1$ for pure geometrical reasons. In fact, $\mathcal P^2+\mathcal V^2$ must equal the square of the length of the Bloch vector and the relation is then saturated by pure states.}
\end{center}
\end{figure}

The freedom to indirectly prepare a state is quite different for a non-selective measurement, i.e. one that does not allow one to wait until the desired result appears. In this case, it is no longer possible to obtain any desired conditional state. At this point we need to introduce the concepts of visibility and predictability. We present here only the necessary introduction to these concepts, for an in-depth description we refer to Bergou and Englert's work \cite{Englert2000}.
Consider a POM, composed of a number $N$ of probability operators $\hat \pi_k$, each corresponding to one of the possible outcomes of a measurement on the environment. We recall that these operators are hermitian, positive, they sum to the identity, but need not be mutually orthogonal. To each measurement outcome corresponds a conditional state of the qubit:
\begin{align}
\hat\rho_k=\frac{\mathrm{Tr_E}[(\mathbb{\hat 1}\otimes\hat\pi_k)\hat\rho]}{p_k},
\end{align}
where $p_k=\mathrm{Tr}[(\mathbb{\hat 1}\otimes\hat\pi_k)\hat\rho]$ is the probability of the $k$-th outcome and the partial trace is calculated over the environment. This state is at some location in the Bloch sphere, at a distance $\mathcal V_k$ from the N-S line, and at a distance $\mathcal P_k$ from the equatorial plane, see Fig.~\ref{definitions}. In the language of the Bloch vector $\mathbf{r}=(x,y,z)$, i.e. if one writes the state in the form $\hat\rho=\frac{1}{2}(\mathbb{\hat 1}+\mathbf{r}\cdot\boldsymbol{\hat\sigma})$, one could write $\mathcal{V}=\sqrt{x^2+y^2}$ and $\mathcal{P}=|z|$. These two distances are called ``visibility'' and ``predictability'' and clearly depend on the choice of basis: if we considered two different opposite points on the surface as the new North and South poles, the distance of the state $\hat\rho_k$ from the new N-S line and the new equatorial plane would change (the only exception being for the maximally mixed state at the center of the sphere, for which $\mathcal V=\mathcal P=0$ regardless of the choice of basis). The visibility $\mathcal V_k$ is a measure of the degree of coherence of the two alternatives that define the North and South poles. The predictability $\mathcal P_k$ is a measure of our ability to predict which of the two will occur upon a measurement of the qubit in that specific basis. Our POM identifies $N$ conditional qubit states, each of which displays its own visibility and predictability. One can then calculate the statistical average of these quantities:
\begin{align}
\bar{\mathcal V}&=\sum_kp_k\mathcal {V}_k=\sum_k \left|\mathrm{Tr}\left[((\hat\sigma_x+i\hat\sigma_y)\otimes\hat\pi_k)\hat\rho\right]\right|\\
\bar{\mathcal P}&=\sum_kp_k\mathcal {P}_k=\sum_k \left|\mathrm{Tr}\left[(\hat \sigma_z\otimes\hat\pi_k)\hat\rho\right]\right|,
\end{align}
where the sums run from 1 to $N$ and where the absolute value of $\hat\sigma_x+i\hat\sigma_y$ measures the distance from the N-S line and the absolute value of $\hat\sigma_z$ measures the distance from the equatorial plane.  We stress that $\bar{\mathcal V}$ and $\bar{\mathcal P}$ are not the expectation values of some operators, because of the absolute value which wraps the trace. There is also a deeper reason why there is no observable which corresponds to these quantities, and it is that if it existed, one could violate the no-signalling principle.

It is very simple to prove that the values of $\bar{\mathcal V}$ and $\bar{\mathcal P}$ that a POM allows us to infer on the qubit are going to be greater or equal than those obtained by ignoring the environment:
\begin{align}
\bar{\mathcal P}&=\sum_k \left|\mathrm{Tr}\left[(\hat \sigma_z\otimes\hat\pi_k)\hat\rho\right]\right|\geq\left| \sum_k\mathrm{Tr}\left[(\hat \sigma_z\otimes\hat\pi_k)\hat\rho\right]\right|=\mathcal P,
\end{align}
and analogously for the visibility. Here we used the fact that $\sum_k\hat\pi_k=\mathbb{\hat 1}$. Therefore, $\mathcal V$ is the lower bound of the average visibility and it is achieved when ignoring the environment. Similarly, $\mathcal P$ is the lower bound of the average predictability and it is achieved when ignoring the environment. What about the upper bounds? One defines the coherence $\mathcal C\leq1$ as the upper bound of the average visibility and the distinguishability $\mathcal D\leq1$ as the upper bound the of average predictability, which are achieved by employing the optimal POMs on the \emph{whole} environment: not having access to the whole environment will inevitably hinder the possibility of reaching $\mathcal C$ and $\mathcal D$. Lastly, note that in general,  the POM that maximizes $\bar{\mathcal P}$ does not automatically maximize $\bar{\mathcal V}$ and vice versa. With this in mind, we can write the following hierarchies:
\begin{subequations}
\begin{align}
\mathcal P&\leq\bar{\mathcal P}\leq\mathcal D,\\
\mathcal V&\leq\bar{\mathcal V}\leq\mathcal C.
\end{align}
\end{subequations}

The standard duality relation $\mathcal P^2+\mathcal V^2\leq 1$ can be justified geometrically by interpreting $\mathcal P$ and $\mathcal V$ as in Fig.~\ref{definitions}. It contains the lower bounds of predictability and visibility, and therefore it refers to a situation in which the environment is not taken into account. If the environment is measured, one has to replace those lower bounds with the averages: $\bar{\mathcal P}^2+\bar{\mathcal V}^2\leq 1$. If one implements a which-alternative measurement, $\bar{\mathcal P}$ will reach the distinguishability, and one obtains ${\mathcal D}^2+\bar{\mathcal V}^2\leq 1$. Complementarily, if one implements an erasure measurement, $\bar{\mathcal V}$ will reach the coherence, and one obtains $\bar{\mathcal P}^2+{\mathcal C}^2\leq 1$. However, we note that as in general these two optimal measurements differ, the quantity ${\mathcal D}^2+{\mathcal C}^2$ can exceed the value of 1.

Therefore with a non-selective measurement one obtains an ensemble of conditional states whose values of $\bar{\mathcal P}$ and $\bar{\mathcal V}$ are limited by the bounds given above. This explains why one does not have the freedom to indirectly prepare the qubit in any desired state. In contrast, we saw that in case selective measurements were allowed, one would eventually (given nonzero correlations between qubit and environment) obtain a state anywhere on or in the Bloch sphere.

\section{Optimal erasure bound}

\begin{figure}[!t]
\begin{center}
\includegraphics[width=0.75\columnwidth]{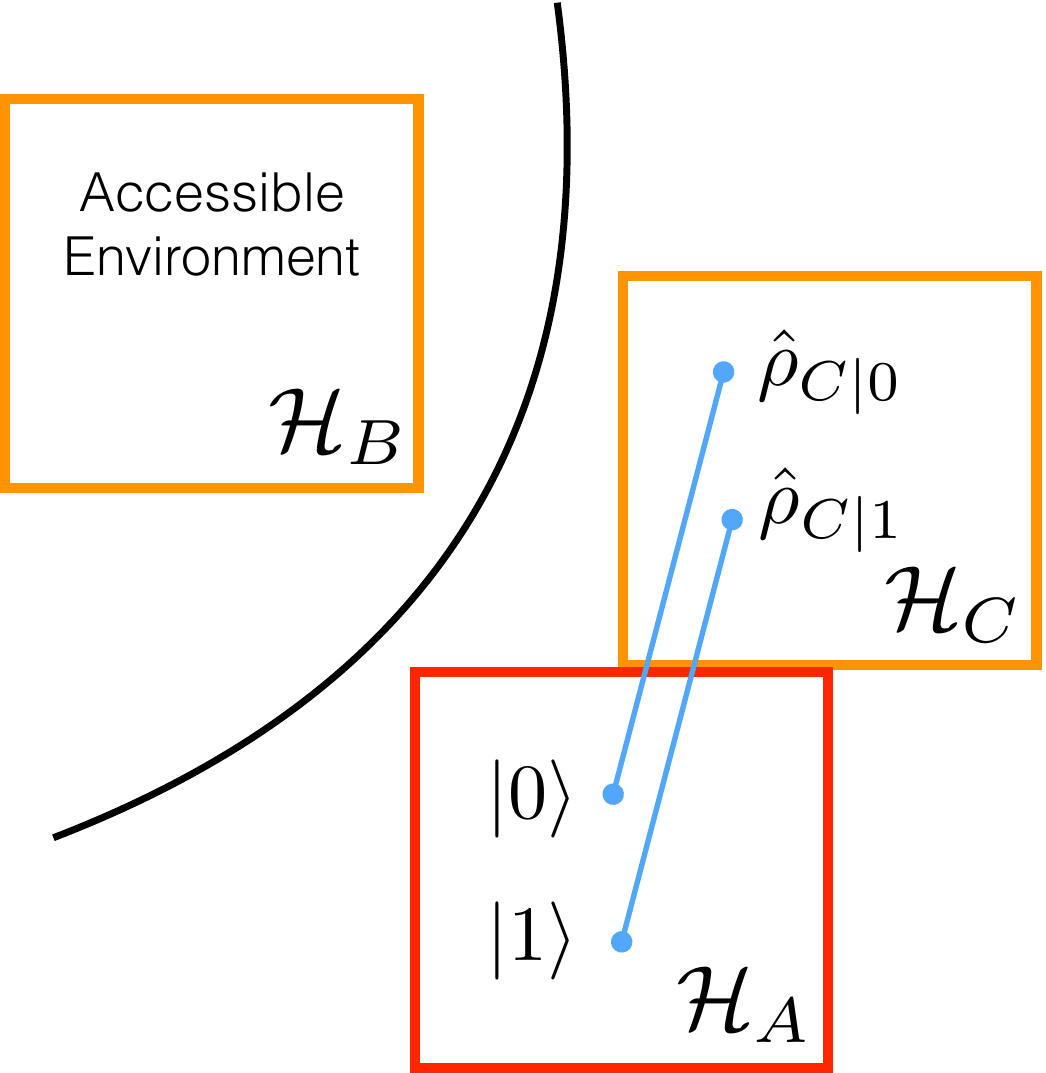}
\caption{\label{fig1} The conditional states of the inaccessible environment that determine the optimal erasure bound are conditioned solely on the eigenstates of the $A$ qubit and (being inaccessible) cannot be conditioned on $B$.}
\end{center}
\end{figure}

Let's now consider the situation described in Fig.~\ref{fig1}. We are facing the task of erasing the information about the alternatives of $A$ that is stored in $B$, by projecting $B$ in the most appropriate basis. We are looking at how well we can perform this task, and how much the state of $C$ matters.

We start by considering the purification $|\rho_{ABC}\rangle$ of the qubit plus the environment.
After we fix the computational basis on the Bloch sphere of $A$, we can write the (unnormalized) conditional states of $B$ and $C$ as
\begin{align}
\tilde \rho_{B|k}&=\mathrm{Tr}_{AC}[(|k\rangle\langle k|\otimes\mathbb{\hat 1}_B\otimes \mathbb{\hat 1}_C)\hat \rho_{ABC}],\\
\tilde \rho_{C|k}&=\mathrm{Tr}_{AB}[(|k\rangle\langle k|\otimes\mathbb{\hat 1}_B\otimes \mathbb{\hat 1}_C)\hat \rho_{ABC}],
\end{align}
where the vertical bar notation is intended to mean ``given the qubit $A$ in the computational state $k=0,1$'' and $|k\rangle\langle k|$ is the projector on the computational states of $A$. We use a tilde to remind that the state is unnormalized, to normalize it we would have to divide it by the probability of measuring the projector $|k\rangle\langle k|$, i.e. $\hat \rho_{C|k}=\tilde \rho_{C|k}/\mathrm{Tr}(\tilde \rho_{C|k})$. Using unnormalized states simplifies the equations below, so we will postpone normalization factors to the end. Note that the state of $A$ and $B$ is given by the density matrix
\begin{align}
\hat\rho_{AB}=\begin{pmatrix}
\tilde\rho_{B|0}&\tilde\chi_B\\
\tilde\chi_B^\dagger&\tilde\rho_{B|1}
\end{pmatrix}
\end{align}
From this matrix, we need two operators: the unnormalized off-diagonal block and the unnormalized difference between the diagonal blocks
\begin{align}
\tilde \chi_B=\mathrm{Tr}_{AC}[(|1\rangle\langle0|\otimes\mathbb{\hat 1}_B\otimes \mathbb{\hat 1}_C)\hat \rho_{ABC}&],\nonumber\\
\tilde\rho_{B|0}-\tilde\rho_{B|1}=\mathrm{Tr}_{AC}[(\hat\sigma_z\otimes\mathbb{\hat 1}_B\otimes \mathbb{\hat 1}_C)\hat \rho_{ABC}&].
\end{align}
We now have all we need to define the key quantity that we want to calculate (the largest visibility of $A$ that can be retrieved by optimizing a quantum erasure POM on $B$) and the largest predictability of the alternatives of $A$ that can be retrieved by optimizing a which-alternative POM on $B$ (which we deal with in the appendix):

\begin{align}
\label{PVDC}
\mathcal C_{A|B}&=\sup_{\mathrm{POM}_B}\sum_kp_k\mathcal {V}_k=2\,\mathrm{Tr}\left|\tilde\chi_B\right|\\
\mathcal D_{A|B}&=\sup_{\mathrm{POM}_B}\sum_kp_k\mathcal {P}_k=\mathrm{Tr}\left|\tilde \rho_{B|0}-\tilde \rho_{B|1}\right|.
\end{align}
The trace norm $\mathrm{Tr}|x|$ corresponds to the sum of the singular values of $x$, which are the eigenvalues of the positive matrix $|x|=\sqrt{x^\dagger x}$.
Note that we used a notation that resembles the coherence because $\mathcal C_{A|B}$ is the highest value of $\bar{\mathcal V}$ that can be retrieved on the qubit $A$ by accessing only $B$ (hence the subscripts). Had we the ability to access the whole environment, the value that we would achieve would be the true coherence $\mathcal C$.
In addressing this problem, we ask the question of how large $\bar{\mathcal V}$ can be, given the constraints imposed by correlations and measurements. As we said, we restrict the measurements to those that span $\mathcal H_B$, so $\bar{\mathcal V}\leq\mathcal C_{A|B}$ and we now calculate this upper bound.

At this point we use the assumption that $\mathrm{dim}(\mathcal H_B)=2$, so $\tilde{\chi}_B^\dagger\tilde{\chi}_B$ will have two positive eigenvalues. Call them $a$ and $b$, it holds that $\mathrm{Tr}|\tilde\chi_B|=\sqrt{a}+\sqrt{b}$. We can express the sum of two square roots in terms of the elementary symmetric polynomials in two variables $s_1=a+b$ and $s_2=ab$:
\begin{align}
\label{key}
\mathrm{Tr}|\tilde\chi_B|=\sqrt{a}+\sqrt{b}=\sqrt{s_1+2\sqrt{s_2}}
\end{align}
The last thing to do is to express the symmetric polynomials in terms of traces, which can be done elegantly via Newton's identities:
\begin{subequations}
\label{newton}
\begin{align}
s_1&=\mathrm{Tr}(x)\\
2s_2&=\mathrm{Tr}(x)^2-\mathrm{Tr}(x^2)\\
6s_3&=\mathrm{Tr}(x)^3-3\mathrm{Tr}(x)\mathrm{Tr}(x^2)+2\mathrm{Tr}(x^3)\\
\dots\nonumber
\end{align}
\end{subequations}
In our case $x=\tilde\chi_B^\dagger \tilde\chi_B$.
After a bit of algebra (see appendix) we find
\begin{align}
\mathrm{Tr}|\tilde\chi_B|^2=E(\tilde \rho_{C|0},\tilde \rho_{C|1})
\end{align}
where $E(\tilde \rho_{C|0},\tilde \rho_{C|1})$ is the sub-fidelity of $\tilde \rho_{C|0}$ and $\tilde \rho_{C|1}$. The sub-fidelity is a lower bound of Uhlmann's fidelity $F(x,y)=\mathrm{Tr}(\sqrt{\sqrt{x}\,y\sqrt{x}})$ and is defined as
\begin{align}
E(x,y)=\mathrm{Tr}(xy)+\sqrt{2}\sqrt{\mathrm{Tr}(xy)^2-\mathrm{Tr}(xyxy)}.
\end{align}
This allows us to write the bound $\mathcal C_{A|B}$ as
\begin{align}
\label{result}
\mathcal C_{A|B}=2\sqrt{E(\tilde \rho_{C|0},\tilde \rho_{C|1})}=2\sqrt{p_0p_1E(\hat \rho_{C|0},\hat \rho_{C|1})}
\end{align}
Where we re-introduced the normalization of the states and exploited the bilinearity of sub-fidelity.
We have therefore found a fundamental link between sub-fidelity and the highest visibility achievable in quantum erasure with minimal access to the environment. Interestingly, under some conditions one can turn the argument around and \emph{define} the sub-fidelity of two states of a system of arbitrary dimension, as the highest visibility that can be reached by acting on one of two qubits that are correlated to it. This would also allow indirect measurements of the sub-fidelity of inaccessible states.

Following similar steps we can extend our analysis to the case $\mathrm{dim}(\mathcal H_B)=3$, i.e. the case where one can access a three-dimensional subspace of the environment. In this case, some simple algebra will tell us that
\begin{align}
\mathrm{Tr}|\tilde\chi_B|&=\sqrt{a}+\sqrt{b}+\sqrt{c}\\
&=\sqrt{s_1+2\sqrt{s_2+2\sqrt{s_3}\,\mathrm{Tr}|\tilde\chi_B|}}
\label{dim3}
\end{align}
where now the symmetric polynomials are in three variables: $s_1=a+b+c$, $s_2=ab+bc+ca$ and $s_3=abc$ and they still satisfy Eq.~\eqref{newton}. So one can solve Eq.~\eqref{dim3} for $\mathrm{Tr}|\tilde\chi_B|$ and still find the highest visibility analytically. It is in principle possible to extend this method to higher dimensions, but it becomes quickly intractable because the number of terms grows exponentially.

\section{average bound}

We now turn our attention to a very interesting problem: we want to find the \emph{average} performance of optimal quantum erasure, i.e. we want to compare the ``raw'' visibility of a random qubit with the visibility after performing optimal erasure on its environment. We can calculate the former analytically, and we will compare it with a numerical evaluation of the latter, making the observation that the ratio between the two is practically independent of the size of the environment.

Technically, we need to find the average of $\mathcal C_{A|B}$ over random states in $B$ with respect to the measure that is induced by tracing away a $2K$-dimensional environment (i.e. the 2-dimensional space $\mathcal H_B$ and a $K$-dimensional space $\mathcal H_C$).
One (slow) way to do this would be to uniformly generate random pure states in the whole space $\mathcal H_A\otimes\mathcal H_B\otimes \mathcal H_C$, then trace $\mathcal H_B$ away, find the operators $\tilde\rho_{C|0}$ and $\tilde\rho_{C|1}$ and calculate their sub-fidelity.
A much quicker way to do this is to generate random states directly through complex random gaussian matrices, which is a quite remarkable method: generate an $m\times n$ matrix $\mu$, with entries sampled from the gaussian distribution in the complex plane centered on the origin and with unit variance. Then, all $n\times n$ density matrices $\rho=\mu^\dagger\mu/\mathrm{Tr}(\mu^\dagger\mu)$ are distributed according to the induced trace measure obtained from tracing $m$ dimensions away from an $mn$-dimensional Hilbert space from which we are sampling uniformly \cite{KarolBook}. In our case $n=2$ and $m=2K$. We will perform this task numerically for environments of dimension $K$ up to $10^3$ within reasonable computation time.

\begin{figure}[h!]
\begin{center}
\includegraphics[width=1.\columnwidth]{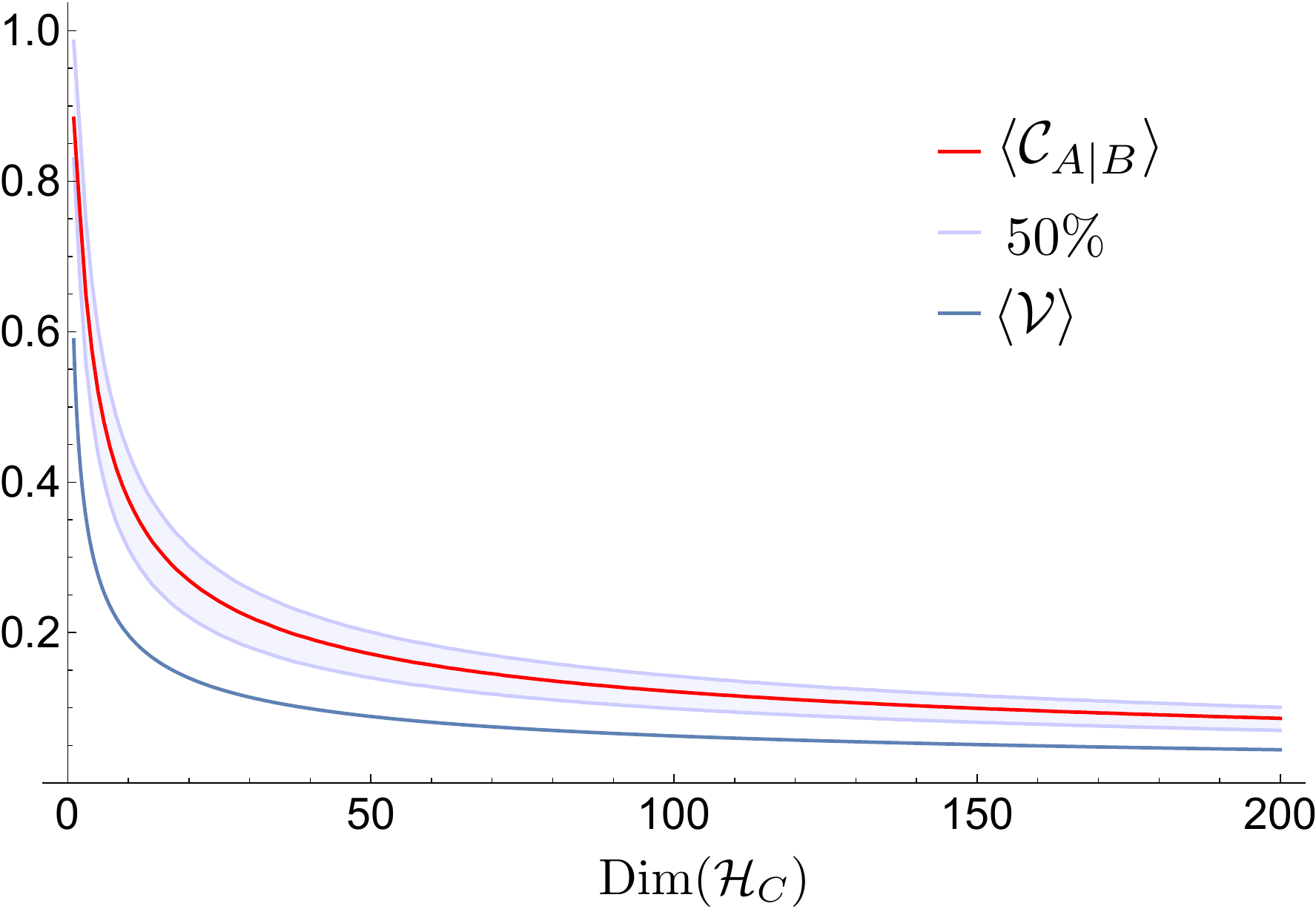}
\caption{\label{fig3} The \emph{average} coherence (Red) that can be retrieved on qubit $A$ by acting on qubit $B$ is about twice the average coherence (Blue) of the qubit $A$ alone (Eq.~\eqref{avgVis}), and it scales asymptotically like the inverse square root of the dimension of the inaccessible environment $\mathrm{Dim}(\mathcal H_C)$. Each plot point is the result of averaging over 200\,000 random states, sampled with an induced partial-trace measure. The bands show the 50\% percentile range around the mean.}
\end{center}
\end{figure}

\begin{figure}[h!]
\begin{center}
\includegraphics[width=1.\columnwidth]{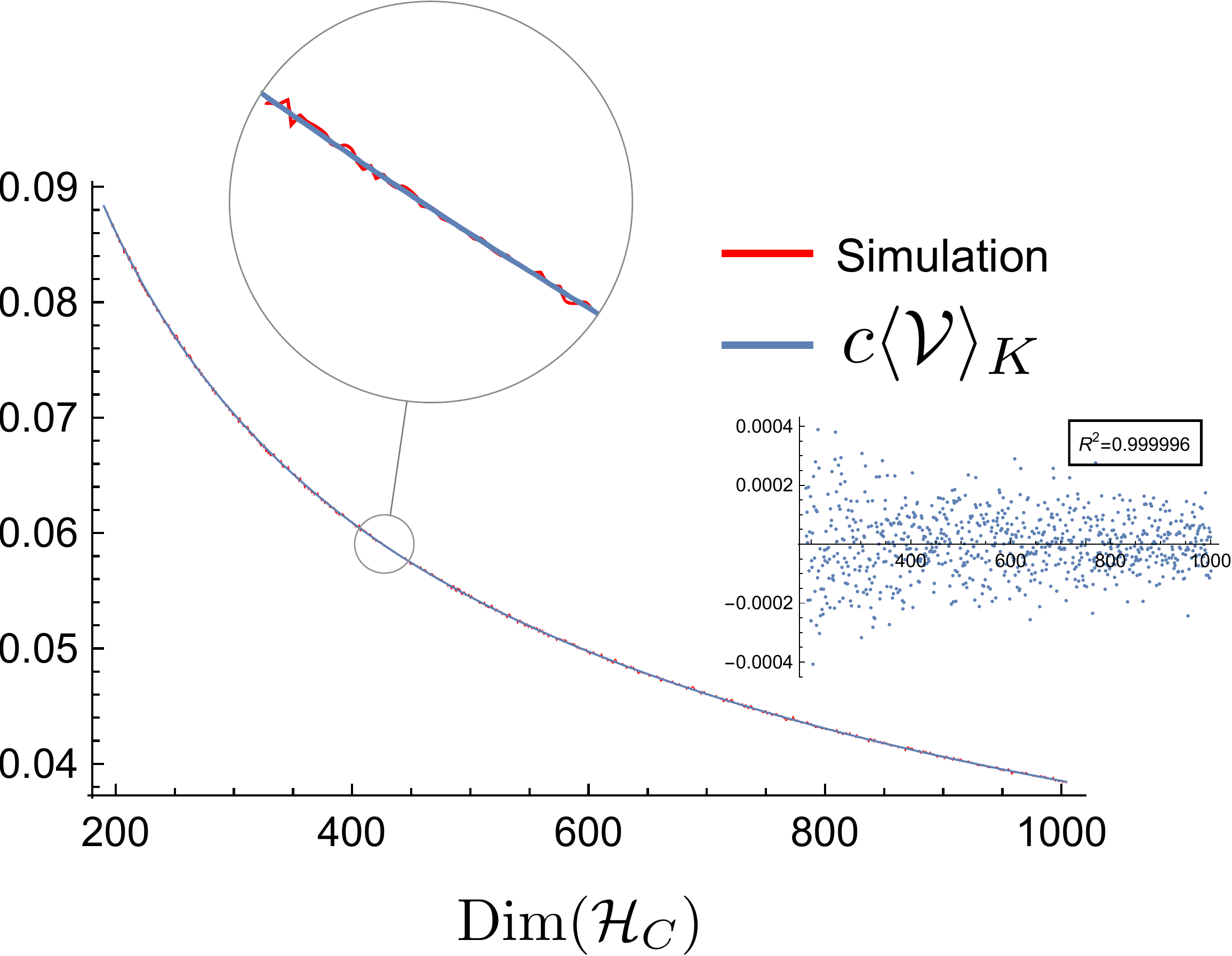}
\caption{\label{fit} Simulation with $200\leq\mathrm{Dim}(\mathcal H_C)\leq1000$, where each point is the average over 20\,000 random states. The quality of the fit is remarkable (on the right, the residues), which shows that for $K$ not too small, $\langle C\rangle_K\sim c\langle V\rangle_K$, where we estimated $c=1.94382\pm0.00013$. This means that one can expect to almost double the visibility of a qubit by performing quantum erasure on a 2-dimensional subspace of its environment.}
\end{center}
\end{figure}
Interestingly, we can still find the average bound analytically for $K=1$, i.e. in the case of an environment entirely constituted by a 2-dimensional accessible space $\mathcal H_B$. In this case, $\mathcal C_{A|B}=2\sqrt{p_0p_1}$, and we can calculate the average of this quantity by sampling mixed states from all the Bloch ball of the qubit with a uniform measure. This cannot be done if $\mathrm{Dim}(\mathcal H_C)>1$, in which case the measure will be more biased towards the center of the Bloch ball. If we indicate with $z$ the vertical coordinate with origin at the centre of the sphere, we have $p_0=(1+z)/2$, $p_1=(1-z)/2$ and the uniform measure on the sphere is $\frac{3}{4} (1 - z^2)\,dz$, so the result is
\begin{align}
\langle\mathcal C_{A|B}\rangle&=\int_{-1}^1 2 \sqrt{\frac{1+z}{2}\frac{1 - z}{2}}\, \frac{3}{4} (1 - z^2)\,dz\nonumber\\
&=\frac{9\pi}{32}\approx0.88357\dots
\end{align}
We note that this result is quite remarkable on its own right: what is says is that given a known random state of two qubits, \emph{on average} one could prepare a single qubit state with an average visibility of almost 90\%. This figure is destined to decrease as the dimension of $\mathcal H_C$ grows, so we are interested in understanding how quickly it does so.
As we are interested in comparing this scaling with the raw visibility of the qubit alone, without any intervention in the space $\mathcal H_B$, we need to calculate $\langle\mathcal V\rangle$. We can do so analytically. We start from the eigenvalue distribution induced by the partial trace $P^\mathrm{trace}_{2,K}(\lambda)$, where $K$ is the dimension of the environment (which for us is going to be $2\mathrm{Dim(\mathcal H_C)}$, where the factor 2 is coming from the dimension of $\mathcal H_B$). For a qubit state, one eigenvalue is sufficient, as the other is determined by the fact that the trace of the density matrix has to be 1. We know from \cite{KarolBook} that 
\begin{align}
P^\mathrm{trace}_{2,K}(\lambda)=\frac{\Gamma(2 K)}{2 \Gamma(K) \Gamma(K - 1)} (\lambda - \lambda^2)^{K-2} (2 \lambda - 1)^2
\end{align}
Therefore, given a diagonalized state with eigenvalues $\lambda$ and $1-\lambda$, we simply have to apply a uniform random rotation in $SU(2)$ and extract the off-diagonal element $\mu=(1-2 \lambda ) \sin (\theta ) (\cos (\psi )+i \sin (\psi ) \cos (\phi )) (\cos (\theta )+i \sin
   (\theta ) \sin (\psi ) \sin (\phi ))$, written in 4-dimensional polar coordinates (considering $S^3$ as the manifold underlying $SU(2)$). So in summary, we average the visibility $2|\mu|$ over $SU(2)$ with the uniform Haar measure $dg$ and over the eigenvalue space with the induced trace measure to obtain:
\begin{align}
\langle \mathcal V\rangle_K&=\int_0^1d\lambda \int_{SU(2)}dg\,2|\mu|P^\mathrm{trace}_{2,K}(\lambda)\nonumber\\
&=\frac{\pi}{4^{K}}\frac{\Gamma(2K)}{K\Gamma(K)^2}
\label{avgVis}
\end{align}
Which in the limit for large $K$, scales like $O(K^{-1/2})$. Recalling that in our case $K=2\mathrm{Dim}(\mathcal H_C)$, one readily obtains the blue curve in Fig.~\ref{fig3}. In case of a pure random two-qubit state (i.e. if there are no correlations with any environment), one obtains the value $\langle\mathcal V\rangle_2=3\pi/16\approx0.58905$.  How does $\langle \mathcal C_{A|B}\rangle_K$ compare with $\langle \mathcal V\rangle_K$? In other words, what is the advantage of performing quantum erasure? We find that the advantage does not depend on the dimension of $\mathcal H_C$. In fact, as the dimensionality of the environment increases, the value of the average coherence becomes a \emph{constant} multiple of the average visibility, i.e. 
\begin{align}
\langle \mathcal C_{A|B}\rangle_K\sim c\langle \mathcal V\rangle_K\qquad(K\rightarrow\infty).
\end{align}
Although this seems to imply that for small $K$ this relation is not in good health, it actually has an error of less than 2\% already from $K=10$.
We ran a simulation and estimated $c=1.94382\pm0.00013$ to a very high degree of confidence (see Fig.~\ref{fit}). This means that if you were to pluck a random pure state of 2 qubits embedded an inaccessible environment, you can expect to almost double the coherence of one of the qubits by optimally measuring the other.

\section{Conclusion}
In this work we have addressed the limitations of quantum erasure on a qubit when we have minimal access to its environment. We find that the highest visibility of the qubit is proportional to the sub-fidelity of the conditional states of the inaccessible part of the environment. This result provides an operational interpretation of sub-fidelity, an insight into correlated systems and it can also give us a way of measuring the sub-fidelity of inaccessible states. Finally, we found that optimal quantum erasure can almost double the visibility of a random qubit embedded in an arbitrarily large environment of which we control only a 2-dimensional subspace.

\section{Acknowledgements}
F.M.M. thanks Steve Barnett, Gerardo Adesso and Karol \.Zyczkowski for helpful comments.
This work was supported by the Canada Excellence Research Chairs (CERC) Program, the Natural Sciences and Engineering Research Council of Canada (NSERC) and the UK EPSRC.

\section{Appendix}
We now provide our derivation of $\mathcal C_{A|B}$. We first expand $\hat\chi_B=\tilde\chi_B/\sqrt{p_0p_1}$ in its most general form:
\begin{align}
\hat\chi_B=\left( 
\begin{array}{c@{}c}
 \sqrt{r_0s_0}\langle c_{10}|c_{00}\rangle & e^{i\theta'}\sqrt{r_0s_1}\langle c_{11}|c_{00}\rangle\\
 e^{-i\theta}\sqrt{r_1s_0}\langle c_{10}|c_{01}\rangle & e^{-i(\theta-\theta')}\sqrt{r_1s_1}\langle c_{11}|c_{01}\rangle\\
\end{array}\right),
\end{align}
where $|c_{ab}\rangle$ are the states of $C$ conditioned on the alternatives of $A$ and $B$ (being $\mathcal H_B$ 2-dimensional, $B$ is a qubit too). The positive numbers $r_b$ and $s_b$ are the relative probabilities of $|c_{0b}\rangle$ and $|c_{1b}\rangle$, respectively. $\theta$ and $\theta'$ are the phases of the states of $B$ conditioned on the alternatives of $A$. For simplicity, we will rewrite this as
\begin{align}
\hat\chi_B=\begin{pmatrix}
 \alpha & \gamma\\
 \delta & \beta\\
\end{pmatrix}.
\end{align}
Plugging this into Eq.~\eqref{key} gives us
\begin{align}
\mathrm{Tr}|\hat\chi_B|^2=|\alpha|^2+|\beta|^2+|\gamma|^2+|\delta|^2+2|\alpha\beta-\gamma\delta|.
\end{align}
Expanding $|\alpha|^2+|\beta|^2+|\gamma|^2+|\delta|^2$ we obtain
\begin{align}
\mathrm{Tr}(\tilde\rho_{C|00}\tilde\rho_{C|10})&+\mathrm{Tr}(\tilde\rho_{C|00}\tilde\rho_{C|11})+\nonumber\\\mathrm{Tr}(\tilde\rho_{C|01}\tilde\rho_{C|10})&+\mathrm{Tr}(\tilde\rho_{C|01}\tilde\rho_{C|11})\nonumber\\=\mathrm{Tr}(\hat\rho_{C|0}\hat\rho_{C|1})
\end{align}
where $\tilde\rho_{C|0b}=r_b|c_{0b}\rangle\langle c_{0b}|$ and $\tilde\rho_{C|1b}=s_b|c_{1b}\rangle\langle c_{1b}|$ are \emph{unnormalized} states. Consequently, $\hat\rho_{C|a}=\tilde\rho_{C|a0}+\tilde\rho_{C|a1}$ are the normalized states of $C$ conditioned on $A$ while ignoring (tracing away) $B$.
The final term is not as straightforward. We begin first by rewriting $|\alpha\beta-\gamma\delta|$ as $\sqrt{(\alpha\beta-\gamma\delta)(\alpha^*\beta^*-\gamma^*\delta^*)}$. We expand what is under the square root and then add and subtract to it the following term:
\begin{align}
\mathrm{Tr}(\tilde\rho_{C|00}\tilde\rho_{C|11})&\mathrm{Tr}(\tilde\rho_{C|01}\tilde\rho_{C|11})+\nonumber\\
\mathrm{Tr}(\tilde\rho_{C|00}\tilde\rho_{C|10})&\mathrm{Tr}(\tilde\rho_{C|01}\tilde\rho_{C|10}).
\end{align}
We then simplify the result with the identity $\mathrm{Tr}(XY)\mathrm{Tr}(XZ)=\mathrm{Tr}(XYXZ)$, which holds for $X$ rank-1. We obtain
\begin{align}
\mathrm{Tr}|\hat\chi_B|^2&=\mathrm{Tr}(\hat\rho_{C|0}\hat\rho_{C|1})\nonumber\\+&\sqrt{2}\sqrt{[\mathrm{Tr}(\hat\rho_{C|0}\hat\rho_{C|1})]^2-\mathrm{Tr}[(\hat\rho_{C|0}\hat\rho_{C|1})^2]}\nonumber\\
&=E(\hat\rho_{C|0},\hat\rho_{C|1}).
\end{align}
For completeness, we now look at the dual problem of optimizing a which-alternative sorting, i.e. the goal is to maximize the which-alternative information. Again, we are restricted in our measurements to those that span $\mathcal H_B$. We can still use Eq. \eqref{key}, only now we have $x=\tilde \rho_{B|0}-\tilde \rho_{B|1}$, which is hermitian. The hermiticity of $x$ allows us to simplify Eq.~\eqref{key} to 
\begin{align}
\left(\mathrm{Tr}|x|\right)^2=
\begin{cases}
2\mathrm{Tr}(x^2)-\mathrm{Tr}(x)^2\quad&\mathrm{Tr}(x^2)\geq\mathrm{Tr}(x)^2\\
\mathrm{Tr}(x)^2&\mathrm{Tr}(x^2)\leq\mathrm{Tr}(x)^2
\end{cases}.
\end{align}
The first case implies that the accessible space $\mathcal H_B$ contains which-alternative information and we can access it.
The second case is trivial and implies that $\mathcal D_{A|B}^2$ reaches its minimum of $\mathcal P^2=(p_0-p_1)^2$, i.e. the accessible space $\mathcal H_B$ does not carry which-alternative information. So it is not surprising that the upper bound $\mathcal D_{A|B}^2$ depends on the conditional states of $\mathcal H_B$:
\begin{align}
\mathcal D_{A|B}^2=
\begin{cases}
2\mathrm{Tr}[(\tilde \rho_{B|0}-\tilde \rho_{B|1})^2]-\mathcal P^2\\
\mathcal P^2
\end{cases}\label{Db}
\end{align}

\bibliography{Manuscript}
\bibstyle{unsrt}

\end{document}